\documentclass[amsmath,amssymb,prd]{revtex4}
\usepackage{graphicx}

\newcommand{\beq}{\begin{equation}}
\newcommand{\eeq}{\end{equation}}
\newcommand{\bea}{\begin{eqnarray}}
\newcommand{\eea}{\end{eqnarray}}
\newcommand{\bse}{\begin{subequations}}
\newcommand{\ese}{\end{subequations}}

\usepackage{amsmath}
\usepackage{amsfonts}
\usepackage{amssymb}

\newcommand{\bM}{\begin{pmatrix}}
\newcommand{\eM}{\end{pmatrix}}

\usepackage{slashed}

\begin{document}

\title{Testing Lorentz and CPT invariance with neutrinos}
\author{Jorge S. D\'iaz}
\affiliation{Institute for Theoretical Physics, Karlsruhe Institute of Technology (KIT), 76128 Karlsruhe, Germany}

\begin{abstract}
Neutrino experiments can be considered sensitive tools to test Lorentz and CPT invariance.
Taking advantage of the great variety of neutrino experiments, including neutrino oscillations, weak decays, and astrophysical neutrinos, the generic experimental signatures of the breakdown of these fundamental symmetries in the neutrino sector are presented.
\end{abstract}

\maketitle


\section{Introduction}

Since the discovery of parity (P) violation in weak interactions, the potential breakdown of other combinations of discrete symmetries has been an attractive method to search for new physical phenomena.
Additionally, the experimental fact that we live in a matter-dominated Universe has been for decades an indication of certain differences between matter and antimatter.
Today we know that not only parity but also the combined transformation of parity and charge conjugation (CP) is only an approximate symmetry of Nature.
Nonetheless, the amount of CP violation in the quark sector seems insufficient to explain our observed universe, which has led to a search for sources of CP violation in other sectors of the Standard Model (SM).
Additionally, it is tempting to go one step forward and consider the possibility that the combination of parity, charge conjugation, and time-reversal (T) could also be just an approximate symmetry; 
nonetheless, 
CPT invariance arises from some of the basic constituents of local, point-particle quantum field theory.
This indicates that if we want consider CPT violation then at least one of the conditions of our current theories ought to be abandoned.
In order to preserve unitarity and microcausality, 
the least drastic requirement that we can abandon is Lorentz invariance \cite{Greenberg}.
Even though this is the approach followed in the rest of this article,
it is important to mention that other proposals exist in which CPT violation can is implemented,
for example, 
by breaking locality instead of Lorentz invariance \cite{CPT01,CPT04,CPT06,CPT10,CPT11}.

The phenomenology of a possible anisotropy of space and its experimental consequences have been considered since several decades \cite{LNI1,LNI2,LNIexp1,LNIexp2}.
Regarding the anisotropy of spacetime,
a systematic approach to incorporate general terms that break Lorentz symmetry in a theory at the action level can be constructed using effective field theory.
This general framework called Standard-Model Extension (SME) \cite{SME1a,SME1b} incorporates Lorentz-violating terms in the action by contracting operators of conventional fields in the SM with controlling coefficients to preserve coordinate invariance.
A subset of Lorentz-violating operators also break CPT invariance \cite{Greenberg}.
Moreover, writing the SM fields in a curved background allows the incorporation of Lorentz violation in gravity \cite{SME2}.
The SME permits the identification of the relevant observable effects in a variety of experiments, which has triggered a worldwide experimental program to search for these possible deviations from exact Lorentz invariance \cite{tables}. 

Neutrinos, some of the most abundant particles in the Universe, have always surprised us.
These elusive particles participate in a variety of phenomena, from low-energy radioactivity to energetic processes in astrophysics, which give them the potential to serve as interesting probes of fundamental physics.
The neutrino sector of the SME offers a rich phenomenology to be studied with current technologies in a wide range of energies and experimental setups.
As mentioned earlier,
different approaches for CPT-violating physics can be implemented with and without Lorentz invariance \cite{CPT01,CPT04,CPT06,CPT10,CPT11,CPT02,CPT03,CPT08}.
Since local field theories that break CPT always appear with Lorentz violation \cite{Greenberg}, in what follows, the key signatures of deviations of Lorentz and CPT invariance are presented for neutrinos in the context of the SME.

\section{Lorentz-violating neutrinos}
\label{Sec_LVnus}

In the SME,
the neutrino sector is described by the Lagrangian density \cite{LVnu,KM2012}
\begin{equation}\label{L_nu}
\mathcal{L} = \frac{1}{2} \overline\Psi\big(i \gamma^\alpha \partial_\alpha- M + \hat{\mathcal{Q}}\big)\Psi + \text{h.c.},
\end{equation}
where neutrinos and their charge conjugates are represented by the multiplet $\Psi=(\nu_e,\nu_\mu,\nu_\tau,\nu_{e}^C,\nu_{\mu}^C,\nu_{\tau}^C)^T$,
the mass matrix $M$ is generated by a conventional seesaw mechanism,
and $\hat{\mathcal{Q}}$ represents a generic operator for Lorentz violation.
In a basis of Dirac matrices,
this operator can be decomposed in the form \cite{KM2012}
\begin{eqnarray}\label{Q}
\hat{\mathcal{Q}} &=& \hat{\mathcal{S}} + i\hat{\mathcal{P}} \gamma_5 + \hat{\mathcal{V}}^\alpha \gamma_\alpha 
    + \hat{\mathcal{A}}^\alpha \gamma_5\gamma_\alpha  + \frac{1}{2} \hat{\mathcal{T}}^{\alpha\beta} \sigma_{\alpha\beta}.
\end{eqnarray}
In the expansion for $\hat{\mathcal{Q}}$,
each term is a $6\times6$ matrix in generation space that can be decomposed into $3\times3$ Dirac and Majorana  blocks \cite{KM2012}.
From the Lagrangian density \eqref{L_nu}, 
an effective Hamiltonian describing three active neutrinos and three active antineutrinos can be constructed.
At leading order in the Lorentz-violating terms \eqref{Q},
the scalar $\hat{\mathcal{S}}$ and pseudoscalar $\hat{\mathcal{P}}$ components are unobservable \cite{KM2012}.
The propagation and mixing of left-handed neutrinos and right-handed antineutrinos is independently modified by the Dirac component of the vector $\hat{\mathcal{V}}^{\alpha}$ and axial vector $\hat{\mathcal{A}}^{\alpha}$ terms;
whereas the Majorana component of the tensor term $\hat{\mathcal{T}}^{\alpha\beta}$ introduces mixing between neutrinos and antineutrinos.

In the full $6\times6$ Hamiltonian incorporating a conventional Lorentz-invariant part and the Lorentz-violating piece,
a $3\times3$ diagonal block $h_{ab}$ describes the three active neutrinos, whereas the other $3\times3$ diagonal block $h_{\bar a\bar b}$ describes the three active antineutrinos.
The remaining $3\times3$ off-diagonal block $h_{a\bar b}$ corresponds to mixing between neutrinos and antineutrinos arising due to Lorentz violation.
The indices $a,b=e,\mu,\tau$ and $\bar a,\bar b=\bar e,\bar\mu,\bar\tau$ denote the flavors of active neutrinos and antineutrinos, respectively.
Each of the $3\times3$ blocks can be explicitly written in the form \cite{LVnu}
\begin{eqnarray}\label{h1}
h_{ab} &=& |\pmb{p}|\delta_{ab} +\frac{m^2_{ab}}{2|\pmb{p}|} + (a_L)^\alpha_{ab} \hat p_\alpha - (c_L)^{\alpha\beta}_{ab}\hat p_\alpha\hat p_\beta|\pmb{p}|, \\
\label{h2}
h_{\bar a\bar b} &=& |\pmb{p}|\delta_{\bar a\bar b} +\frac{m^2_{\bar a\bar b}}{2|\pmb{p}|} + (a_R)^\alpha_{\bar a\bar b} \hat p_\alpha - (c_R)^{\alpha\beta}_{\bar a\bar b}\hat p_\alpha\hat p_\beta|\pmb{p}|, \\
\label{h3}
h_{a\bar b} &=& i\sqrt{2}(\epsilon_+)_\alpha \big(\tilde{H}^{\alpha}_{a\bar b} - \tilde{g}^{\alpha\beta}_{a\bar b}\hat p_\beta|\pmb{p}| \big), 
\end{eqnarray}
where at leading order the neutrino momentum is given by the energy $|\pmb{p}|\approx E$.
The mass-squared matrix is conventionally written in terms of the Pontecorvo-Maki-Nakagawa-Sakata (PMNS) matrix \cite{PMNS1,PMNS2} and its eigenvalue differences lead to conventional description of neutrino oscillations.
The mass matrix for antineutrinos is related to that of neutrinos by complex conjugation $m^2_{\bar a\bar b} =(m^2_{ab})^*$.
The left- and right-handed coefficients for CPT-odd Lorentz violation are related as $(a_R)^\alpha_{\bar a\bar b}=-(a_L)^{\alpha*}_{ab}$,
while the corresponding coefficients for CPT-even Lorentz violation are related as $(c_R)^{\alpha\beta}_{\bar a\bar b}=(c_L)^{\alpha\beta*}_{ab}$.
In the block Hamiltonian,
the coefficients for Lorentz violation are coupled to the neutrino four momentum $\hat p^\alpha=(1;\hat{\pmb{p}})$ and polarization $(\epsilon_+)^\alpha$ \cite{DKM}.
The incorporation of operators of arbitrary dimension in the theory leads to higher powers of the neutrino energy in the Hamiltonian blocks \eqref{h1}, \eqref{h2}, and \eqref{h3} \cite{KM2012,KM2013}.

The coefficients for Lorentz violation $(a_L)^\alpha_{ab}$ and $(c_L)^{\alpha\beta}_{ab}$ act like fixed background fields that trigger observable effects when the neutrino experiment is rotated and boosted.
For experiments with detector and source on the surface of the Earth, the neutrino beam rotates with sidereal frequency $\omega_\oplus\simeq2\pi/(\text{23 h 56 min})$ with respect to the background fields. 
This time dependence can be used to parametrize the relevant observable in terms of harmonics of the sidereal angle $\omega_\oplus T_\oplus$.
A systematic comparison between different experiments searching for Lorentz-violation requires a common inertial frame; for this reason, experimental results are conventionally reported in the Sun-centered equatorial frame \cite{tables}.
In addition to searching for generic effects of Lorentz violation in neutrinos, some studies have considered the possibility of using the Hamiltonian \eqref{h1} to describe the neutrino-oscillation data as an alternative to the conventional mass-driven oscillations.
These global models based on the SME have interesting features that could accommodate most of the established data as well as some anomalous results reported by different experiments \cite{LVmodels1,LVmodels1b,LVmodels2,LVmodels3,LVmodels4,LVmodels5,LVmodels7}.

The generic signatures of Lorentz and CPT violation that could be observed in neutrino experiments can be classified into two types: oscillation-free effects and neutrino mixing.
The former refers to modifications that affect the behavior of the three neutrino flavors in the same form, which leaves oscillations unaffected because the modifications correspond to factors proportional to the identity matrix in flavor space.
These oscillation-free effects can be observed in kinematical measurements such as neutrino velocity and weak decays. 
On the other hand, 
mixing effects arise due to coefficients that mix the neutrino flavors.
The following sections describe the key signatures of Lorentz and CPT violation that can be studied in neutrino experiments.

\section{Oscillation-free signals}

Lorentz-violating effects that modify the three neutrino flavors by the same amount are unobservable in oscillations because they lead to no mixing.
In this case, one way to study these effects is by considering other kinematical studies, such as the neutrino velocity.
In the presence of Lorentz violation, the deviation of the neutrino velocity from the speed of light is \cite{KM2012}
\begin{eqnarray}\label{v}
v_\nu-1 &=&  \frac{|\mathrm{m}|^2}{2|\pmb{p}|^2} +
\sum_{djm} (d-3)|\pmb{p}|^{d-4}\,e^{im\omega_\oplus T_\oplus}\,_0\mathcal{N}_{jm}(\hat{\pmb{p}}) 
\big((a_\text{of}^{(d)})_{jm}-(c_\text{of}^{(d)})_{jm}\big),
\end{eqnarray}
where the real mass parameter $|\mathrm{m}|^2$ does not participate in oscillations, and we have used a spherical decomposition for the Lorentz-violating contribution.
The possible energy dependence of the neutrino velocity apears in terms of the magnitude of the neutrino momentum $|\pmb{p}|$.
For details on the relationship between different decompositions of the coefficients see Ref. \cite{KM2012}. 
The sum indicates that there are several coefficients that produce independent effects.
Operators of arbitrary dimension in the theory are characterized by the index $d$, whereas the spherical properties of the oscillation-free coefficients for CPT-odd $(a_\text{of}^{(d)})_{jm}$ and CPT-even  $(c_\text{of}^{(d)})_{jm}$ Lorentz violation are labeled by the angular momentum indices $jm$.
The angular factors $_0\mathcal{N}_{jm}(\hat{\pmb{p}})$ contain all the information regarding the orientation of the neutrino beam in the Sun-centered frame in terms of the location of the laboratory and spherical harmonics $Y_{jm}(\hat{\pmb{p}})$ in the laboratory frame. 
The index $m$ also controls the harmonics of the sidereal phase $\omega_\oplus T_\oplus$;
this time dependence is a consequence of the change of the neutrino-beam orientation in the Sun-centered frame due to Earth's rotation \cite{KM2012}.
The use of a spherical basis and the angular momentum indices $jm$ allows identifying the transformation properties under rotations with ease.
For instance,
expression \eqref{v} above shows that the neutrino velocity is isotropic only for $j=0$ because the associated spherical harmonic is $Y_{00}$. 
Similarly,
the neutrino velocity becomes direction dependent for $j\neq0$ and time dependent for $m\neq0$. 
The last two features arise due to the loss of invariance under rotations,
reflected in the coupling with spherical harmonics with $j\geq1$.
Additionally, for odd $d$, CPT violation makes neutrinos and antineutrinos move at different speed (for antineutrinos the coefficient $(a_\text{of}^{(d)})_{jm}$ has opposite sign).
The expression for the neutrino velocity \eqref{v} can be directly applied to this type of study in beam experiments \cite{FNAL76,FNAL79,MINOS_v,OPERA,ICARUS,Borexino,LVD}.
Notice that for operators of dimension $d=3$ there is no modification to the neutrino velocity,
whereas for $d=4$ this modification is independent of the energy $E\approx|\pmb{p}|$.
For $d>4$ in the theory, 
the neutrino velocity \eqref{v} becomes energy dependent, which can produce novel effects including the dispersion of neutrinos of different energies.
Dispersive effects and time delays due to modifications of the neutrino speed can be tested to high precision using distant sources of neutrinos because the long distances traveled enhance the minute effects of the coefficients for Lorentz violation \cite{KM2012,Longo_SN}.

The modified neutrino (and antineutrino) dispersion relation indicates that the coefficients for Lorentz violation can also be studied by searching for modifications in energy-momentum conditions in different processes. 
In other words, the Lorentz-violating contribution can open the phase space of conventionally forbidden processes as well as blocking certain processes that would be otherwise allowed.
Take for instance the leptonic decay of a charged meson $M^+$ of the form $M^+\to l^++\nu_l$, which becomes forbidden above some threshold energy $E_\text{th}$ \cite{KM2012,threshold1,ChodosTh,threshold2,threshold3,threshold4,threshold5,threshold6,threshold8}.
The observation of the decay products at a given energy $E_0$ can then be used to establish the threshold condition $E_\text{th}>E_0$.
Using the Hamiltonian \eqref{h1} in spherical basis and denoting the mass of the relevant charged meson by $M_M$, the threshold condition can be generically written in the form  \cite{KM2012}
\begin{equation}\label{Threshold}
\sum_{djm} E_0^{d-2}\,Y_{jm}(\hat{\pmb{p}}) \big[\pm(a_\text{of}^{(d)})_{jm}-(c_\text{of}^{(d)})_{jm}\big] <
\frac{1}{2}(M_M-m_{l^\pm})^2,
\end{equation}
where $m_{l^\pm}$ indicates the mass of the charged lepton and the sign $+$ ($-$) refers to neutrinos (antineutrinos).
The spherical harmonics $Y_{jm}(\hat{\pmb{p}})$ encode direction-dependent effects.

Clearly the higher the neutrino energy, the better the constraint on the coefficients for Lorentz violation.
The observation of energetic neutrinos in IceCube \cite{IceCube_PeVnus,IceCube_28nus} serves as a very sensitive probe of Lorentz and CPT invariance using this threshold analysis. 
In particular, the observation of several events can be used to go beyond conventional isotropic studies and to explore direction-dependent effects, which requires a spread of events in the sky \cite{VHEnus}.

Following a similar approach, forbidden processes can become allowed due to Lorentz and CPT violation.
For instance, some coefficients in the neutrino velocity \eqref{v} can produce superluminal neutrinos, which could lose energy in the form of Cherenkov radiation \cite{KM2012,ColemanGlashow,Cerenkov1,Cerenkov2,Cerenkov3,Cerenkov4,Cerenkov5,Cerenkov6,Cerenkov7,Cerenkov8,Cerenkov9,
Cerenkov11}.
The determination of a characteristic energy-loss distance $D(E)=-E/(dE/dx)$ due to Cherenkov radiation can be used to write the condition $L<D(E)$, where $L$ is the distance traveled by the neutrino \cite{KM2012}.
For the electron-positron emission in the process $\nu\to\nu+e^-+e^+$ the rate of energy loss takes the explicit form \cite{KM2012,VHEnus}
\begin{equation}
\frac{dE}{dx}= -
\frac{C}{8}\int \frac{\kappa^0\,\pmb{\kappa}'^2}{(\kappa^2-M_Z^2)^2}\frac{\partial|\pmb{\kappa}'|}{\partial\kappa_0}
\frac{q\cdot k\;q'\!\cdot k'}{q_0k_0q_0'k_0'}\,d^3p'\,d\Omega_{\kappa'},
\end{equation}
where $\kappa=k+k'$ and $\kappa'=k-k'$ are auxiliary 4-vectors defined in terms of the momentum of the electron ($k$) and the positron ($k'$), $C$ is a constant, and $q/q_0=(1,\hat{\pmb{p}})$, $q'/q_0'=(1,\hat{\pmb{p}}')$ encode the Lorentz-violating effects in the neutrino dispersion relation.
This formula can be used to search for Lorentz violation through the observation of high-energy astrophysical neutrinos \cite{VHEnus,Borriello2013,Stecker2013,Stecker2014,Stecker:2015,Mazon2014}.

Despite the high sensitivity to the effects of Lorentz and CPT violation described above, the experimental signatures of operators of dimension $d=3$ are unobservable in the techniques that involve modifications to the neutrino velocity.
For these particular operators in the theory, weak decays are the appropriate experimental setup \cite{DKL}.
For single beta decay, the observable signatures of Lorentz and CPT violation appear due to the modified dispersion relation of the antineutrino as well as the altered spinor solutions of the equation of motion \cite{Diaz_BD}.
Near the endpoint of the beta spectrum, the study of tritium decay using MAC-E filters offers the possibility of searching for anisotropic effects.
The magnetic fields that guide the emitted electrons to be analyzed in the main spectrometer only select decay products within a given acceptance cone, which defines a direction that allows the study of direction-dependent effects.
Location and orientation of the experiment become relevant and fit parameters such as the endpoint energy can depend on sidereal time \cite{DKL,Diaz_BD}.
This idea can be tested in the near future in the KATRIN experiment \cite{KATRIN}. 

Another method to search for an anisotropy of space is by asymmetries in the decay of polarized and unpolarized neutrons in which the direction of the emitted antineutrino can be inferred.
These asymmetries would depend on the location of the experiment and could also vary with sidereal time.
The particular coefficient that produces isotropic effects can be studied by searching for modifications in the full beta spectrum of neutron decay \cite{DKL,Diaz_BD}.
A similar distortion of the full spectrum appears in the two-neutrino mode of double beta decay experiments \cite{Diaz_DBD}.
This alteration of the electron-sum spectrum in two-neutrino double beta decay has been explored by the EXO-200 experiment,
obtaining the first experimental limit on the relevant coefficient for CPT-odd Lorentz violation \cite{EXO:2016}.
We remark in passing that other coefficients for Lorentz and CPT violation can produce other novel effects.
For instance, CPT-violating Majorana couplings in the SME can trigger neutrinoless double beta decay even for a negligible Majorana mass \cite{Diaz_DBD}.

\section{Neutrino oscillations}

Neutrino-oscillation experiments are sensitive interferometers to study the coefficients in the Hamiltonian \eqref{h1} that produce neutrino mixing.
Depending on the details of the experiment of interest, the relevant oscillation probabilities can be calculated using perturbative methods.
In short-baseline accelerator-based experiments the ratio between the baseline $L$ (a few hundreds of meters) and energy $E$ (a few MeV) is too small to produce oscillations, according to the conventional massive-neutrino model because; 
therefore, 
any oscillation signal would be a consequence of Lorentz violation.
Direct calculation shows that for appearance experiments the oscillation probability takes the form \cite{KM_SB}
\begin{equation}\label{P_SB}
P_{\nu_b\to\nu_a}\simeq L^2\big|(a_L)^\alpha_{ab} \hat p_\alpha - (c_L)^{\alpha\beta}_{ab}\hat p_\alpha\hat p_\beta E\big|^2,\quad a\neq b,
\end{equation}
where $a,b$ are the neutrino flavors.
It should be noticed that off-diagonal elements of the Hamiltonian in flavor space are in general complex numbers.
Moreover, the corresponding Hamiltonian for antineutrinos is obtained by replacing $(a_L)^\alpha_{ab}\to-(a_L)^{\alpha*}_{ab}$ and $(c_L)^{\alpha\beta}_{ab}\to(c_L)^{\alpha\beta*}_{ab}$.

On the other hand, long-baseline experiments are designed to have a far detector where the first oscillation maximum occurs according to the massive-neutrino model.
In this case, a large oscillation signal is expected and Lorentz violation can be treated as a small correction to the mass-driven oscillations.
The oscillation probabilities can be written as a power series \cite{DKM}
\begin{equation}\label{P012}
P_{\nu_b\to\nu_a} = P_{\nu_b\to\nu_a}^{(0)} + P_{\nu_b\to\nu_a}^{(1)} + P_{\nu_b\to\nu_a}^{(2)} + \ldots,
\end{equation}
where the zeroth-order term in Lorentz violation $P_{\nu_b\to\nu_a}^{(0)}$ corresponds to the conventional oscillation probability from the massive-neutrino model.
The elements of the series \eqref{P012} can be calculated order by order.
The neutrino beam determines a particular direction that varies with sidereal phase $\omega_\oplus T_\oplus$ and also introduces direction-dependent effects.
For instance, the sub-leading term in the series \eqref{P012} can be written as
\begin{eqnarray}\label{P1}
\frac{P_{\nu_b\to\nu_a}^{(1)}}{2L} &=& (P_{\mathcal{C}}^{(1)})_{ab} 
+(P_{\mathcal{A}_s}^{(1)})_{ab}\sin{\omega_\oplus T_\oplus}%
+(P_{\mathcal{A}_c}^{(1)})_{ab}\cos{\omega_\oplus T_\oplus} \nonumber\\
&&\quad\quad
+(P_{\mathcal{B}_s}^{(1)})_{ab}\sin{2\omega_\oplus T_\oplus}
+(P_{\mathcal{B}_c}^{(1)})_{ab}\cos{2\omega_\oplus T_\oplus}.
\end{eqnarray}
where the amplitudes of the sidereal harmonics depend on the coefficients for Lorentz violation and properties of the experiment such as the energy of the neutrinos, location of the experiment, and the  orientation of the neutrino beam \cite{KM_SB,DKM}.
A similar expression can be found for antineutrino oscillations.

The formulations described above have been implemented to perform several independent and complementary searches for Lorentz and CPT violation 
by Double Chooz \cite{LV_DC}, IceCube \cite{LV_IceCube}, LSND \cite{LV_LSND}, MiniBooNE \cite{LV_MiniBooNE1,LV_MiniBooNE2}, MINOS \cite{LV_MINOS_ND1,LV_MINOS_ND2,LV_MINOS_FD},
and Super-Kamiokande \cite{LV_SK}.
To date there is no compelling evidence for Lorentz violation and these experimental searches have served to constraint several SME coefficients, which are summarized in Ref. \cite{tables}.
The complementarity of these searches is due to the study of different oscillation channels, which depend on similar coefficients but with different flavor indices.

It was mentioned at the beginning of Sec. \ref{Sec_LVnus} that in the presence of Lorentz violation, neutrinos and antineutrinos can also mix.
The independent set of SME coefficients in the Hamiltonian \eqref{h3} can trigger the oscillation between left-handed neutrinos and right-handed antineutrinos.
These oscillations always exhibit direction dependence and are absent in the first two terms of the series \eqref{P012}.
In other words, the SME coefficients that produce neutrino-antineutrino mixing appear quadratically in the oscillation probability.
For these oscillations, the probability can be written in the same form as the linear term \eqref{P1}. However, being a second-order effect the sidereal decomposition will include up to fourth harmonics \cite{DKM}.
Systematic experimental searches of these neutrino-antineutrino oscillations have been performed using accelerator \cite{RebelMufson} and reactor neutrinos \cite{LV_DC2};
nonetheless,
the most sensitive tests of this type have made use of solar neutrinos \cite{Diaz:2016b}.
The long propagation distance of solar neutrinos offers a vast advantage over any terrestrial experiment.
At high energies,
precise measurements of the flavor ratios of astrophysical neutrinos could serve as one of the most sensitive test of Lorentz symmetry \cite{Arguelles:2015}.

\section{Conclusions}

The breakdown of Lorentz and CPT invariance has been studied both theoretically and experimentally in a wide range of systems.
The neutrino sector offers the possibility to perform several types of tests of the validity of these fundamental symmetries.
From the high precision of low-energy experiments studying weak interactions to the high energy of astrophysical neutrinos traveling long distances before reaching our detectors as well as the great power of neutrino oscillations thanks to their interferometric nature.
Lorentz and CPT invariance have survived the several tests performed using the SME as a general framework; nonetheless, there are many other effects that remain unexplored.

Neutrinos have amazed us since they were proposed by Pauli to save the conservation of energy \cite{Pauli}, today neutrinos offer the opportunity to challenge the cornerstone of modern physics.

\section*{Acknowledgments}
The work of JSD was supported in part by the German Research Foundation (DFG) under Grant No. KL 1103/4-1.


%
\end{document}